\documentclass[prd,preprint]{revtex4}

\usepackage{graphicx}
\usepackage{amsmath}
\usepackage{amssymb}
\usepackage[mathscr]{euscript}
\usepackage[verbose,a4paper]{hyperref}
\usepackage{epsfig}
\usepackage{subfigure}

\newcommand{\bal}{\begin{align}}
\newcommand{\eal}{\end{align}}
\newcommand{\bea}{\begin{eqnarray}}
\newcommand{\eea}{\end{eqnarray}}
\newcommand{\beq}{\begin{equation}}
\newcommand{\eeq}{\end{equation}}
\newcommand{\bean}{\begin{eqnarray*}}
\newcommand{\eean}{\end{eqnarray*}}
\newcommand{\nn}{\nonumber}
\newcommand{\barr}{\begin{array}}
\newcommand{\earr}{\end{array}}

\newcommand{\lsim}{\raise0.3ex\hbox{$\;<$\kern-0.75em\raise-1.1ex\hbox{$\sim\;$}}}
\newcommand{\gsim}{\raise0.3ex\hbox{$\;>$\kern-0.75em\raise-1.1ex\hbox{$\sim\;$}}}

\newcommand{\bpm}{\begin{pmatrix}}
\newcommand{\epm}{\end{pmatrix}}
\newcommand{\abs}[1]{\left| #1 \right|}

\newcommand{\BB}{{|B^0B^0\rangle}}
\newcommand{\BbB}{{|\overline B^0B^0\rangle}}
\newcommand{\BBb}{{|B^0 {\overline B}^0\rangle}}
\newcommand{\BbBb}{{|\overline B^0\overline B^0\rangle}}

\newcommand{\Dm}{\Delta M}
\newcommand{\DG}{\Delta\Gamma}

\newcommand{\refe}[1]{Eq.(\ref{#1})}

\begin{document}

\vspace*{-2cm}
\begin{flushright}
IFIC/04-65\\
FTUV-04-1029\\
October 2004
\end{flushright}

\title{CPT violation in entangled $B^0-\overline B^0$ states \\ 
and the demise of flavour tagging}
\author{E.~Alvarez$^a$}
\author{J.~Bernab\'eu$^a$} 
\author{N.E.~Mavromatos$^b$} 
\author{M.~Nebot$^a$}
\author{J.~Papavassiliou$^a$}
\affiliation{$^a$ Departamento de F\'\i sica Te\'orica and IFIC, Centro Mixto, 
Universidad de Valencia-CSIC,
E-46100, Burjassot, Valencia, Spain. \\
$^b$ King's College London, Department of Physics, Theoretical Physics, 
Strand WC2R 2LS, London, U.K.}

\begin{abstract}

We  discuss the  demise of  flavour  tagging due  to the  loss of  the
particle-antiparticle   identity   of    neutral   B-mesons   in   the
Einstein-Podolsky-Rosen correlated states.  Such a situation occurs in
cases where the CPT operator  is ill-defined, as happens, for example,
in  quantum gravity  models  with induced  decoherence  in the  matter
sector.   The time  evolution of  the perturbed  $B^0-{\overline B}^0$
initial state,  as produced in B-factories, is  sufficient to generate
new two-body states.   For flavour specific decays at  equal times, we
discuss two  definite tests of  the two body entanglement:  (i) search
for  the would-be  forbidden $B^0B^0$  and  ${\overline B}^0{\overline
B}^0$ states;  (ii) deviations from  the indistinguishable probability
between  the permuted states  ${\overline B}^0B^0$  and $B^0{\overline
B}^0$.

\end{abstract}

\maketitle

The determination of the initial 
flavour of a single neutral meson
is usually referred to as 
``flavour tagging'', and 
is a technique employed in a variety of 
experiments~\cite{tagging}. In the case of $\phi$- and $B$-factories,  
where the neutral meson states produced 
( $K^0$--$\overline K^0$ and $B^0$--$\overline B^0$, respectively) constitute  
correlated Einstein-Podolsky-Rosen states (EPR)~\cite{dunietz,botella,bernabeu}, 
the 
knowledge that one of the two mesons decays at a given time 
through a flavour specific channel (``tagging'') allows one to 
unambiguously 
infer  the flavour of   the accompanying  meson state at the same time.  
Thus, for example, 
the detection of a flavour 
specific $B^0$ (or $\overline B^0$) decay on one side of the detector implies 
a pure $\bar B^0$ (or $B^0$) 
state on the other side  (we will always refer to the fra
me associated with the resonance, the center-of-mass frame).
In this article we will argue that the 
basic underlying (and usually unquestioned) assumptions, leading to 
the above conclusion, may be invalidated 
if the CPT operator cannot be {\it intrinsically} defined.
These latter circumstances may occur, for example, in the context of 
an extended class of quantum gravity models, where the structure of quantum space time at 
Planckian scales ($10^{-35}$ m)
may actually be fuzzy, 
characterised by a ``foamy'' nature ({\it space time foam})~\cite{foam}.
In addition, we will propose a set of basic observables, whose measurement would 
effectively amount to a direct testing of the validity of the  
hypothesis associated with the tagging.

In what follows we will go over the assumptions built into the 
flavour tagging with EPR states.  
In the conventional 
formulations of {\it entangled} meson  
states~\cite{dunietz,botella,bernabeu}
one imposes the requirement of {\it Bose statistics} 
for the state $K^0 {\overline K}^0$ or $B^0 {\overline B}^0$. 
This, in turn, implies that the physical neutral meson-antimeson state 
must be {\it symmetric} under the combined operation C${\cal P}$,
where C is the charge conjugation and 
${\cal P}$ the operator that permutes the spatial coordinates. 
Specifically, assuming 
{\it conservation} of angular momentum, and 
a proper existence of the {\it antiparticle state} (denoted by a ``bar''),
one observes that, 
for $K^0 {\overline K}^0$ states which are C-selfconjugates with 
C$=(-1)^\ell$ (with $\ell$ the angular momentum quantum number), 
the system has to be an eigenstate of 
${\cal P}$ with eigenvalue $(-1)^\ell$. 
Hence, for $\ell =1$, we have that C$=-$, implying ${\cal P}=-$.
As a consequence 
of Bose statistics this ensures that for $\ell = 1$ 
the state of two identical bosons is forbidden~\cite{dunietz}.
What is more, the probability of obtaining identical 
decay channels at equal times exactly vanishes, independently 
of CP, T and/or CPT violation in the effective hamiltonian. 
  
As a result, the initial entangled state 
$B^0{\overline B}^0$ produced in a B factory 
can be written as:
{\small 
\begin{equation} 
|\psi(0)\rangle = \frac{1}{\sqrt{2}}\left(|B^0({\vec k}),{\overline B}^0(-{\vec k})\rangle
- |{\overline B}^0({\vec k}),{B}^0(-{\vec k})\rangle\right),
\label{bbar}
\end{equation}}
where B-meson momenta are $\pm \vec k$ and $\vec k\cdot \vec p_{e^-}>0$, with $\vec p_{e^-}$ the momentum of the colliding $e^-$.
This specific form of the state vector
is intimately connected with 
the procedure of tagging. In particular, the   
 antisymmetric nature of the state 
under permutation forbids the presence of $\BB$ and $\BbBb$ terms.
It is elementary to verify that, under normal Hamiltonian 
evolution of the system, this latter property persists: at any given time the state 
remains  antisymmetric, given by 
\beq
|\psi(t)\rangle=\frac{e^{-iMt-\frac{\Gamma}{2}t}}{\sqrt{2}}
\left\{ \BBb - \BbB \right\} .
\label{eq:stateT}
\eeq

Evidently, detection of a given  flavour at any time $t$ implies the presence of the 
opposite flavour at the same time and in opposite sides of the detector.


However, as has been pointed out for the first time in
\cite{Bernabeu:2003ym}, 
the assumptions leading to 
Eq.(\ref{bbar}) may not be valid if CPT symmetry is violated, 
not in the usually considered sense of the CPT operator 
not commuting with the hamiltonian of the system at hand~\cite{cpt}, 
but rather in a way which most likely occurs in quantum gravity. 
Namely, a decoherent quantum evolution takes place 
in the ``medium'' of a space time foam~\cite{foam}, in which case 
pure states evolve into mixed ones, 
a scattering S-matrix cannot be properly defined,  
and hence, according to the 
theorem of Ref.~\cite{wald}, 
the CPT operator is not well defined, 
thereby leading to a
strong form of CPT violation.
In such a case~\cite{Bernabeu:2003ym} 
${\overline B}^0$ cannot be considered 
as identical to ${B}^0$, and thus the requirement of C${\cal P} = +$, imposed 
by Bose statistics, is relaxed.
As a result, the initial entangled state (\ref{bbar}) 
can be parametrised in general as:
\beq
|\psi(0)\rangle=\frac{1}{\sqrt{2\left(1+\abs{\omega}^2\right)}}\left\{\BBb-\BbB+\omega\left[\BBb+\BbB\right]\right\},
\label{initialstate}
\eeq
where $\omega = |\omega| e^{i\Omega}$ 
is a {\it complex} CPT-violating parameter~\cite{Bernabeu:2003ym}, 
associated with the non-identical particle nature 
of the neutral meson
and antimeson states. We emphasize that the modification 
in \refe{initialstate} is due to the 
loss of indistinguishability of $B^0$ and $\overline B^0$ and not due to 
violation of symmetries in the production process.
Evidently, the probabilities 
for the two states connected by 
a permutation are different due to the presence of $\omega$.

This modification of the initial state vector has far-reaching consequences for the 
concept of particle tagging. In what follows we will study the time 
evolution of 
(\ref{initialstate}), in order to (i) establish the appearance of terms of the 
(previously forbidden) type $\BB$ and $\BbBb$, and (ii) introduce a set 
of observables, which could actually serve as a direct way for 
checking 
experimentally the robustness of the correlation between the two states 
assumed during the tagging. 

The eigenstates of the effective hamiltonian 
with well defined time evolution are
\begin{align}
& |B_1\rangle=\frac{1}{\sqrt{2\left(1+\abs{\epsilon_1}^2\right)}}\left((1+\epsilon_1)|B^0\rangle+(1-\epsilon_1)|\overline B^0\rangle\right), \nn\\
& |B_2\rangle=\frac{1}{\sqrt{2\left(1+\abs{\epsilon_2}^2\right)}}\left((1+\epsilon_2)|B^0\rangle-(1-\epsilon_2)|\overline B^0\rangle\right).
\end{align}
The eigenvalues of the effective hamiltonian corresponding to $|B_1\rangle$ and $|B_2\rangle$ are, 
respectively, $\mu_1=M_1+i\Gamma_1/2$ 
and $\mu_2=M_2+i \Gamma_2/2$, and we define the quantities 
$M=(M_1+M_2)/2$,\, 
$\Dm=M_1-M_2$,\, $\Gamma=(\Gamma_1+\Gamma_2)/2$,\, and $\DG=\Gamma_1-\Gamma_2$.

Thus, written in terms of the states $|B_1\rangle$ and $|B_2\rangle$, 
the initial state $|\psi(0)\rangle$ in \refe{initialstate} assumes the form
\beq
|\psi(0)\rangle=\frac{1} {\sqrt{2\left(1+\abs{\omega}^2\right)}}
\Bigg\{C_{12}|B_1B_2\rangle +C_{21}|B_2B_1\rangle +C_{11}|B_1B_1\rangle +C_{22}|B_2B_2\rangle \Bigg\},
\label{eq:MUstate-initial}
\eeq
where 
\begin{align}
C_{12}&= \frac{\sqrt{(1+\abs{\epsilon_1}^2)(1+\abs{\epsilon_2}^2)}}{\epsilon_1\epsilon_2-1}\left(1-\omega\frac{\epsilon_1-\epsilon_2}{\epsilon_1\epsilon_2-1}\right),\nn \\
C_{21}&= - \frac{\sqrt{(1+\abs{\epsilon_1}^2)(1+\abs{\epsilon_2}^2)}}{\epsilon_1\epsilon_2-1}
\left(1+\omega\frac{\epsilon_1-\epsilon_2}{\epsilon_1\epsilon_2-1}\right), \nn \\
C_{11}&= \omega\frac{(1-\epsilon_2^2)(1+\abs{\epsilon_1}^2)}{(1-\epsilon_1\epsilon_2)^2},\nn \\
C_{22}&= -\omega\frac{(1-\epsilon_1^2)(1+\abs{\epsilon_2}^2)}{(1-\epsilon_1\epsilon_2)^2}.\label{eq:MUstate-initial-coef}
\end{align}
We note the presence of $|B_1B_1\rangle$ and $|B_2B_2\rangle$,  
which is a characteristic feature when $\omega \neq 0$. Furthermore, 
$C_{12} \ne -C_{21}$.

With the quantum mechanical effective 
Hamiltonian time evolution, the states at a later time $t$ are given by
\beq
|B_1(0)\rangle \mapsto e^{-iMt-\frac{\Gamma}{2} t}e^{-i\frac{\Dm}{2}t-\frac{\DG}{4}t}|B_1(0)\rangle \,,\,\,\,\,\,\,
|B_2(0)\rangle  \mapsto  e^{-iMt-\frac{\Gamma}{2} t}e^{+i\frac{\Dm}{2}t+\frac{\DG}{4}t}|B_2(0)\rangle \,,
\label{eq:MUstates-time}
\eeq
implying that the various terms in \refe{eq:MUstate-initial} 
will have in general a different time evolution. Returning to the flavour-state basis, 
since we are interested in flavour specific decay channels, we have
\beq
|\psi(t)\rangle=\frac{e^{-iMt-\frac{\Gamma}{2}t}}{\sqrt{2\left(1+\abs{\omega}^2\right)}}\left\{ C_{0\bar 0}(t)\BBb +C_{\bar 0 0}(t)\BbB +C_{00}(t)\BB +C_{\bar 0\bar 0}(t)\BbBb\right\} ,
\label{eq:FLAVOURstate-timesa}
\eeq
where
\begin{align}
C_{0\bar 0}(t)&=1+ \omega f(t) \,,\nn\\
C_{\bar 00}(t)&=-1+ \omega f(t) \,,\nn\\
C_{00}(t)&= \frac{\omega}{\left(1-\epsilon^2+\frac{\delta^2}{4}\right)^2}\left((1+\epsilon)^2-\frac{\delta^2}{4}\right)
\bigg(f_1(t)+f_2(t)\bigg)\,,
\nn\\
C_{\bar 0\bar 0}(t)&=\frac{\omega}
{\left(1-\epsilon^2+\frac{\delta^2}{4}\right)^2}\left((1-\epsilon)^2-\frac{\delta^2}{4}\right)\bigg(f_1(t)-f_2(t)\bigg)\,,
\label{eq:FLAVOURstate-Cs}
\end{align}
with
\bea
f(t) &=& \frac{1}{\left(1-\epsilon^2+\frac{\delta^2}{4}\right)^2}
\left[\delta^2+\frac{1}{2}\left((1+\epsilon)^2-\frac{\delta^2}{4}\right)
\left((1-\epsilon)^2-\frac{\delta^2}{4}\right)\left(e^{\alpha t}+e^{-\alpha t}\right)\right]\,,
\nn\\
f_1(t) &=& -\frac{1}{2}\left(1-\epsilon^2+\frac{\delta^2}{4}\right)
\left(e^{\alpha t}-e^{-\alpha t}\right)\,,
\nn\\
f_2(t) &=& -\delta+\frac{\delta}{2}\left(e^{\alpha t}+e^{-\alpha t}\right)\,,
\label{fdef}
\eea
and we have defined $\epsilon=(\epsilon_1+\epsilon_2)/2$, $\delta=\epsilon_1-\epsilon_2$, and 
$\alpha\equiv i \Dm/2 + \DG/4$. We emphasize that the above expressions are exact; no expansion 
with respect to any of the parameters has taken place. 
Phase redefinitions of the single $B$-meson states such as 
$B^0\mapsto e^{i\gamma}B^0$, $\overline B^0\mapsto e^{i\bar \gamma}\overline B^0$ 
are easily handled through the transformation of $\{\epsilon,\delta\}$-dependent expressions. 
The most useful properties under the above mentioned rephasings are
\[
\frac{\delta}{1-\epsilon^2+\frac{\delta^2}{4}}\mapsto \frac{\delta}{1-\epsilon^2+\frac{\delta^2}{4}}\quad ;\quad \frac{(1\pm \epsilon)^2-\frac{\delta^2}{4}}{1-\epsilon^2+\frac{\delta^2}{4}} \mapsto \frac{(1\pm \epsilon)^2-\frac{\delta^2}{4}}{1-\epsilon^2+\frac{\delta^2}{4}}~e^{\pm i(\gamma-\bar\gamma)}.
\]
They lead to explicitly rephasing invariant $C_{0\bar 0}(t)$ and $C_{\bar 0 0}(t)$ 
coefficients, whereas $C_{00}(t)\mapsto e^{i(\gamma-\bar\gamma)}C_{00}(t)$ 
and $C_{\bar 0\bar 0}(t)\mapsto e^{i(\bar\gamma-\gamma)}C_{\bar 0\bar 0}(t)$ 
are individually rephasing-variant, but their dependence on the phase is such 
that the considered physical observables are rephasing invariant, as they should.  
As a check we note that setting $t=0$ in 
the above expressions for $C_{ab}$ we recover the state of 
Eq.(\ref{initialstate}). 

Evidently, a non-vanishing $\omega$ allows both symmetric and 
antisymmetric terms under $B^0\leftrightarrows \overline B^0$. Thus, 
contrary to the standard $\omega=0$ case where the antisymmetric nature 
of the state forbids the presence of $\BB$ and $\BbBb$ terms, 
both $\BB$ and $\BbBb$ terms appear at $t\neq 0$. 
This result has an important  
consequence on the concept of flavour tagging: 
in the presence of the 
$\omega$ effect, 
the detection of a flavour 
specific $B^0$ (or $\overline B^0$) decay on one side does not necessarily imply 
a pure $\bar B^0$ (or $B^0$) 
state on the other side. Clearly, there is a minute ``contamination'', due to the  
presence on one side of the same meson that has been actually tagged 
at the opposite side.

Having concluded the demise 
of the concept of tagging in the presence of $\omega$, it would be 
interesting to invent a set of observables which would actually measure the deviation, if any, from
the basic tagging assumption. We will focus on 
observables involving {\it simultaneous} $B^0$ or $\overline B^0$ 
flavour specific decays. 
This eliminates the standard terms $B^0(t)B^0(t+\Delta t)$ and $\overline B^0(t)\overline B^0(t+\Delta t)$ as they vanish for $\Delta t=0$.
In what follows we will restrict our attention to the most characteristic 
case of flavour specific channels, 
namely semileptonic decays. The main reason for this choice is the fact 
that the  flavour specificity of such decays relies on a 
minimum number of assumptions, in particular solely on the equality  $\Delta B=\Delta Q$, 
and is completely independent of whether or not the CP and CPT symmetries are exact~\cite{foot}.
We emphasize that 
other flavour specific channels may not share this property 
when there is CP or CPT violation in the decay.
Notice also that any effects stemming from 
the possibly decoherent 
(i.e. non quantum-mechanical) 
evolution of the initial state can be unambiguously separated from the $\omega$ effect 
through the difference in the symmetry properties of their contributions to 
the density matrix \cite{Bernabeu:2003ym}.
  
Our basic observables are equal time intensities of flavour specific decays of $B$ mesons.
We consider the four flavour specific channels $| X_{00}\rangle$, $| X_{\bar 0\bar 0}\rangle$, 
$| X_{0\bar 0}\rangle$ and 
$| X_{\bar 00}\rangle$, characteristic to the $B$-meson combinations $\BB$, $\BbBb$,  $\BBb$, and $\BbB$, respectively.
Since,
$\abs{\langle X_{ab}| B^{c} B^{d}\rangle} \sim \delta_{a}^{c}\delta_{b}^{d} $, 
with $ab,cd$ =$00,\bar 0\bar 0,0\bar 0,\bar 00$ (in this compact notation,   
$| \overline B^0 \rangle \equiv | B^{\bar 0}\rangle$, etc),
 it is evident that 
sandwiching the state of Eq.(\ref{eq:FLAVOURstate-timesa}) 
with one of the aforementioned flavour specific channels projects out the 
corresponding co-factor $C_{ab}$. Defining the four intensities 
$I_{ab}(t) = \abs{\langle X_{ab}|\psi(t)\rangle}^2$ we find that 
\begin{equation}
I_{ab}(t)=\abs{\langle Y_{a}| B^{a}\rangle }^2
\abs{\langle Z_{b}| B^{b}\rangle }^2
\frac{e^{-\Gamma t}}{2(1+\abs{\omega}^2)}\abs{C_{ab}(t)}^2 ,
\label{eq:FLAVOURintensities:01}
\end{equation}
where the state $|X_{ab}\rangle$ has been decomposed into the 
two single-meson flavour-specific decay states, $Y_{a}$ and $Z_{b}$, i.e.
$|X_{ab}\rangle = |Y_{a},Z_{b}\rangle $.
These equal-time intensities can be easily time-integrated:
\beq
{\cal{I}}_{ab}=\int_0^\infty \!\!\!\!\! dt~ I_{ab}(t). 
\label{eq:FLAVOURintensities:02}
\eeq

As seen in Eq.~(\ref{eq:FLAVOURstate-Cs}), the parameters 
involved in the time evolution, $\epsilon, \delta, \Dm, \Delta \Gamma$ 
only appear in terms which are explicitly proportional to $\omega$. 
For the $B^0-{\overline B}^0$ system, the terms 
proportional to $\omega \delta$ and $\omega \Delta \Gamma$ can be 
considered as higher order.

In terms of intensities, $\omega\neq 0$ allows
\[
I_{00}(t)\neq 0\quad;\quad  I_{\bar 0\bar 0}(t)\neq 0~.
\]
It is through these otherwise (for $\omega=0$) forbidden intensities that we can explore the presence of $\omega\neq 0$. 
As we can see in \refe{eq:FLAVOURstate-Cs} and \refe{eq:FLAVOURintensities:01},
 what one hopes to observe  is an $\abs{\omega}^2$ vs. $0$ effect.
This would be an {\it unambiguous} manifestation of our effect, 
independently of any other source of symmetry violation.

In the hypothetical situation of  non-vanishing values for 
$I_{00}(t)$ and  $I_{\bar 0\bar 0}(t)$ one could 
consider a CP-type asymmetry of the form  
\begin{equation}
  A_{CP}(t)=\frac{I_{00}(t)-I_{\bar 0\bar 0}(t)}{I_{00}(t)+I_{\bar 0\bar 0}(t)} \quad ; \quad 
{\cal A}_{CP}=\frac{{\cal{I}}_{00}-{\cal{I}}_{\bar 0\bar 0}}
{{\cal{I}}_{00}+{\cal{I}}_{\bar 0\bar 0}} ~~ . \label{eq:Asymmetries:01}
\end{equation} 
The asymmetries $A_{CP}(t)$ and ${\cal A}_{CP}$ express the difference
between the  decay rates of  $B^0 \rightarrow X_0$ and  $\overline B^0
\rightarrow X_{\bar 0}$, where, as before, $X_0$ is a specific flavour
channel  and  $X_{\bar 0}$  its  C-conjugate  state  (in our  notation
$\overline X_0 \equiv  X_{\bar 0}$).  In order to  isolate the physics
associated with 
$C_{00}$ and $C_{\bar 0\bar 0}$  through an observable such as $A_{CP}(t)$, 
one must eliminate its dependence 
on  the decay amplitudes  
$\abs{\langle Y_{a}| B^{a}\rangle }^2
\abs{\langle Z_{b}| B^{b}\rangle }^2$ 
entering through  \refe{eq:FLAVOURintensities:01}.
If the physics governing the decay is
CPT-invariant  (as in  the  Standard Model),  the  use of  inclusive
channels   guarantees  the   cancellation  of  the decay   amplitudes  in
\refe{eq:Asymmetries:01}.  
If  we   consider  exclusive  channels instead,  CP
violation  in   the  decays  prevents in general the aforementioned 
cancellation from 
taking place, thus restricting the usefulness of $A_{CP}(t)$. 
In addition to these  standard  considerations,
quantum gravity itself may affect  the CPT invariance in
the decays; nevertheless, such contributions will be subleading, 
and we will neglect them in what follows.

Interestingly enough,  $A_{CP}(t)$ and ${\cal A}_{CP}$ are 
independent of the value of $\omega$, since the latter 
clearly cancels out when forming the corresponding ratios, to leading order, when quantum gravity induced CPT violating effects 
in the decays are ignored. 
For $\delta=0$ and $\DG$ small, such that terms of order $\omega \DG$ can be 
safely neglected, 
\refe{eq:Asymmetries:01} simplifies to
\begin{equation}
A_{CP}(t)={\cal A}_{CP}=\frac{\abs{1+\epsilon}^4-\abs{1-\epsilon}^4}{\abs{1+\epsilon}^4+\abs{1-\epsilon}^4}=\frac{4~(1+\abs{\epsilon}^2)~\Re e~ \epsilon}{(1+\abs{\epsilon}^2)^2+(2~\Re e~ \epsilon)^2} ~.
\label{eq:asymm2}
\end{equation}
In terms of the standard mixing parameters $p$ and $q$,
\[
\frac{\abs{1+\epsilon}^4-\abs{1-\epsilon}^4}{\abs{1+\epsilon}^4+\abs{1-\epsilon}^4}=\frac{\abs{p}^4-\abs{q}^4}{\abs{p}^4+\abs{q}^4}=\frac{2\Delta_B}{1+\Delta_B^2}~,
\]
where 
\[
\Delta_B=\frac{2~\Im m(M_{12}^* \Gamma_{12})}{(\Dm)^2+\abs{\Gamma_{12}}^2}.
\]
According to the present measurements of the semileptonic rate asymmetry \cite{ASemileptonic}, $A_{CP}(t)={\cal A}_{CP}=-0.007\pm 0.013$.

The algebraic cancellation of all the $\omega$ 
dependence in \refe{eq:Asymmetries:01} can be 
physically understood by realizing that $\omega\neq 0$ allows the equal time presence of 
$\BB$ and $\BbBb$ terms, and it has nothing to do with $B^0$--$\overline  B^0$ 
 mixing  or  $B^0,{\overline  B}^0$
decays. As mentioned previously, possible quantum 
gravity effects in the decays contribute to higher order (at least linear
in $\omega$-like parameters) 
terms in \refe{eq:asymm2}.
The CP asymmetries in \refe{eq:Asymmetries:01} are thus \emph{conventional} 
CP asymmetries between states which are both CPT-forbidden; 
this cancellation is an explicit proof of both effects. This provides an additional way of testing 
the self-consistency of the entire procedure: Once non-vanishing $I_{00}(t)$ and  $I_{\bar 0\bar 0}(t)$ have been established 
one should extract the experimental value of ${\cal A}_{CP}$, which should coincide with the 
theoretical expression  of \refe{eq:Asymmetries:01}; for 
the calculation of the latter one needs as 
input only the standard value for the parameter $\epsilon$, with no reference 
to the actual value of $\omega$.


To isolate linear effects in $\omega$, we pay attention to 
the channels ``$0\bar 0$'' and ``$\bar 00$'' and consider the following CPT-violating, \emph{exchange asymmetries}:
\begin{equation}
A(t)=\frac{I_{0\bar 0}(t)-I_{\bar 00}(t)}{I_{0\bar 0}(t)+I_{\bar 00}(t)}\quad ; \quad 
\mathcal{A}=\frac{\mathcal I_{0\bar 0}-
\mathcal I_{\bar 00}}{\mathcal I_{0\bar 0}+\mathcal I_{\bar 00}}~~.\label{eq:Asymmetries:03}
\end{equation}
As in $A_{CP}(t)$  and ${\cal A}_{CP}$, we are interested in eliminating, in \refe{eq:Asymmetries:03}, the effects related to the decays:
this is again accomplished through CPT-invariant 
inclusive 
semileptonic decays or CP-conserving flavour specific hadronic channels. 
As we shall explain below, 
$A(t)$ and ${\cal A}$ measure the difference between 
the amplitudes corresponding to the permuted states $\BBb$ and $\BbB$. 

\begin{figure}[htb]
\begin{center}
\subfigure[\label{fig01a}]{\epsfig{file=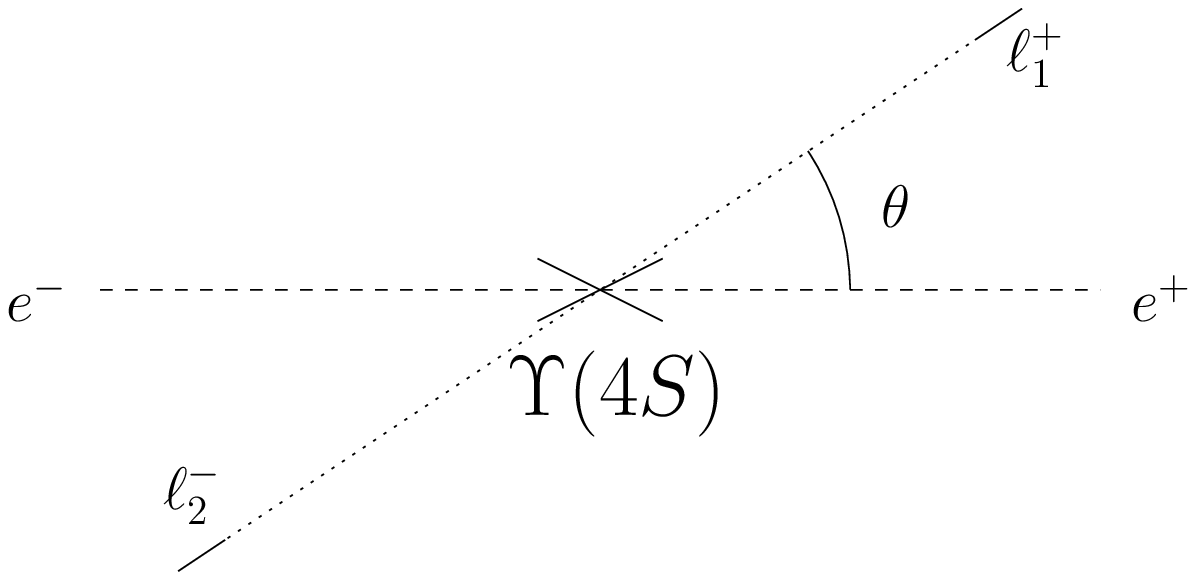,width=0.4\textwidth}}\qquad \subfigure[\label{fig01b}]{\epsfig{file=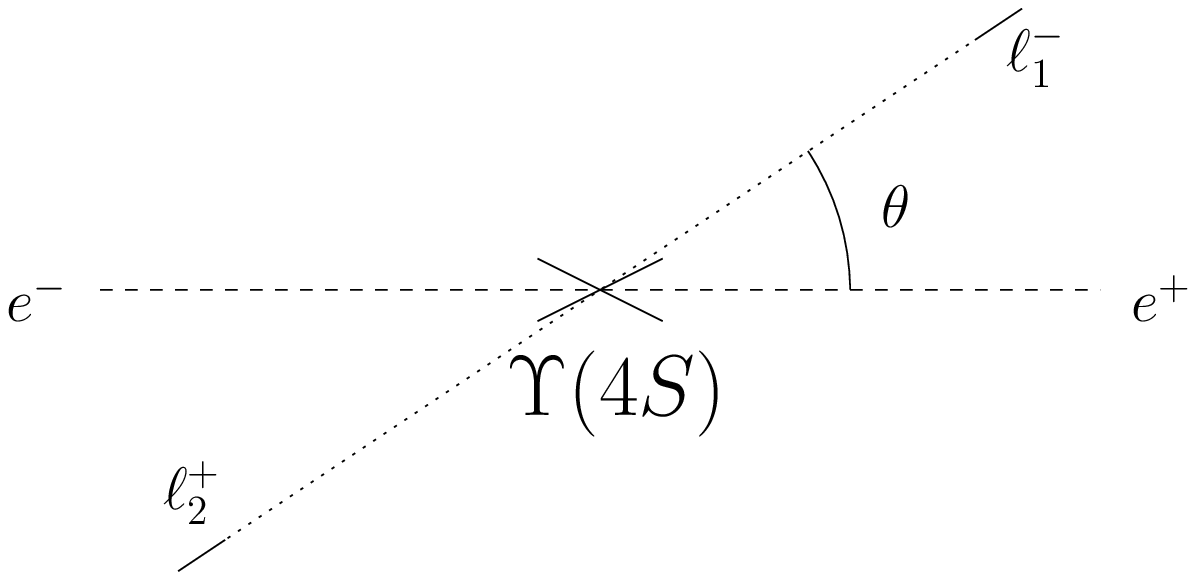,width=0.4\textwidth}}\\
\subfigure[\label{fig01c}]{\epsfig{file=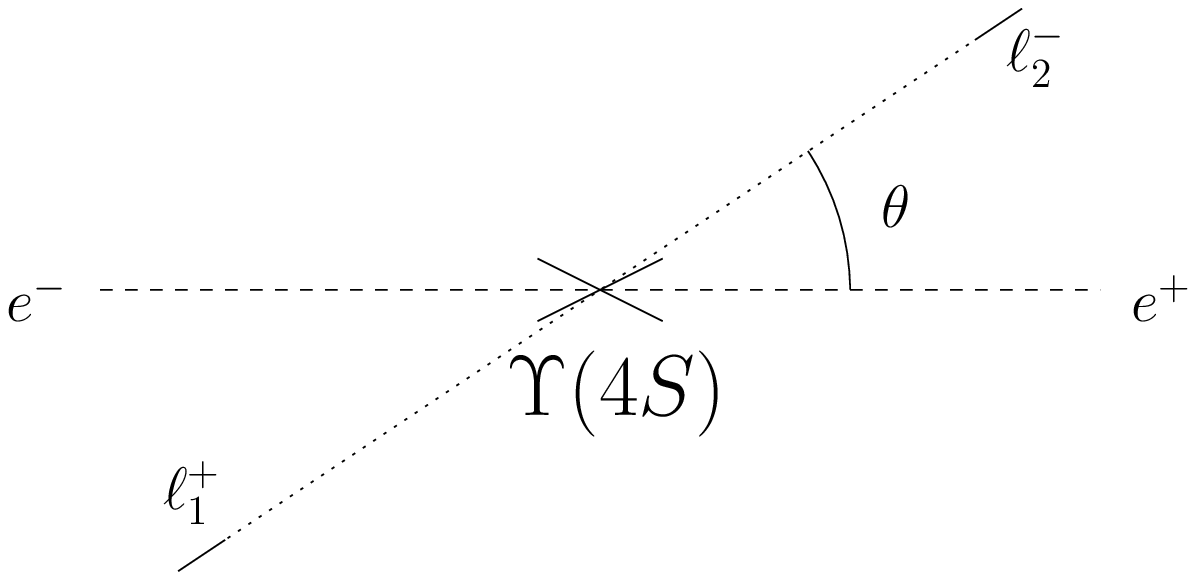,width=0.4\textwidth}}\qquad \subfigure[\label{fig01d}]{\epsfig{file=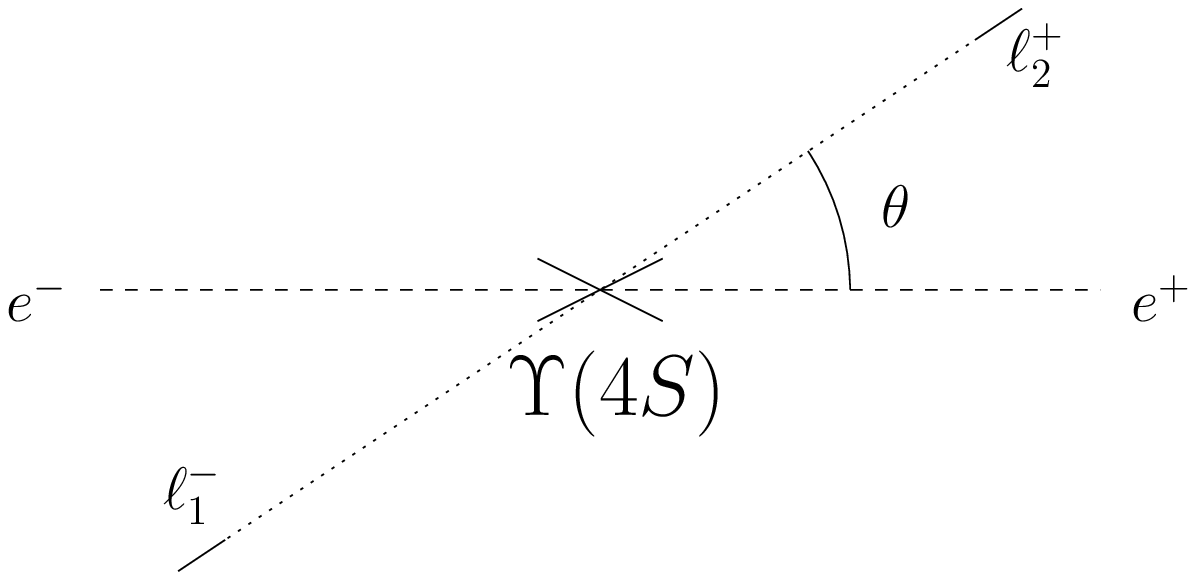,width=0.4\textwidth}}
\end{center}
\caption{Schematic events in inclusive semileptonic B decays. The leptons 
are the only final particles shown, for brevity.}
\label{fig01}
\end{figure}

To this end, we find it instructive to clarify first  
some crucial concepts with the help of figure \ref{fig01},
which depicts inclusive 
semileptonic B decays, for definiteness. For our purposes,
the situation is identical to flavour-specific hadronic channels. 
When $\omega=0$, the states $\BBb$ and $\BbB$ are related through charge conjugation C and through the permutation $B^0\leftrightarrows \overline B^0$; as a consequence of Bose symmetry, 
{\it no observable} can distinguish between those states. 
Notice that 
this fact does not rely on the definition of two-particle states. Indeed, 
recall  that $\BBb$ stands for $|B^0(\vec k) \overline B^0(-\vec k)\rangle$, 
where, as pointed out after \refe{bbar},  $\vec k$ 
is such that $\vec k\cdot \vec p_{e^-}>0$ (this implies 
$0\leq\theta <\frac{\pi}{2}$ for the situation depicted 
in fig.~\ref{fig01}). The schematic events shown in the figure 
correspond unambiguously to the two-particle state 
that is actually projected out:
\[
\ref{fig01a}\to \BBb\quad;\quad\ref{fig01b}\to \BbB\quad;\quad\ref{fig01c}\to \BbB\quad;\quad\ref{fig01d}\to \BBb~.
\]
When $\omega=0$, the identity $I_{0\bar 0}(t)=I_{\bar 00}(t)$ is independent of our $\vec k$-dependent 
two particle convention. The situation changes 
drastically when $\omega\neq 0$. First of all, notice that
the pairs (\ref{fig01a} and \ref{fig01d}) 
and (\ref{fig01b} and \ref{fig01c}) in fig.~\ref{fig01}
are related through charge conjugation C, while 
the pairs (\ref{fig01a} and \ref{fig01c}) 
and (\ref{fig01b} and \ref{fig01d})
are related through permutations $B^0\leftrightarrows \overline B^0$.

As the permutation $B^0\leftrightarrows \overline B^0$ 
is no longer a symmetry, any sensible definition 
of two-particle states 
should not include contributions 
related through the permutation
$B^0\leftrightarrows \overline B^0$ 
in the same intensity $I_{0\bar 0}(t)$ or $I_{\bar 00}(t)$. 
Note also that there is an invariance of these intensities
under rotations around the colliding $e^-e^+$ direction. 
Our $\vec k$-dependent definition is the simplest one that guarantees these 
properties. 
Indeed, 
events of the type \ref{fig01a} and \ref{fig01d} 
contribute to $I_{0\bar 0}(t)$, while 
events like \ref{fig01b} and \ref{fig01c} contribute to $I_{\bar 00}(t)$.
Under charge conjugation, $I_{0\bar 0}(t)\to I_{0\bar 0}(t)$ 
and $I_{\bar 00}(t)\to I_{\bar 00}(t)$, whilst
under $B^0\leftrightarrows \overline B^0$, $I_{0\bar 0}(t)\to I_{\bar 00}(t)$ and $I_{\bar 00}(t)\to I_{0\bar 0}(t)$.
From the above discussion, then, it becomes clear 
that $A(t)$ and ${\cal A}$ measure the asymmetry originated by the permutation $B^0\leftrightarrows \overline B^0$.

Using the expressions 
for $C_{ab}$ given in \refe{eq:FLAVOURstate-Cs}, it is
straightforward 
to establish that $A(t)$  depends, to leading order, linearly on  $\omega$, 
due to the interference  between  the $\omega$-dependent  
and  the standard, $\omega$-independent  (``1''), parts 
of  $C_{0\bar 0}(t)$  and $C_{\bar   00}(t)$:
\begin{equation}
A(t) = \frac{2 \Re e \left(\omega f(t)\right)}
{1+\abs{\omega f(t)}^2}~~.\label{eq:Asymmetries:04}
\end{equation}
where $f(t)$ is defined in (\ref{fdef}).

For $\delta=0$ and $\DG\to 0$, \refe{eq:Asymmetries:04} simplifies to
\begin{equation}
A(t) =\frac{2~\Re e(\omega)~\cos\left(\Dm~ t/2\right)}{1+\abs{\omega}^2\cos^2\left(\Dm ~t/2\right)} \quad ; \quad \mathcal{A}=\frac{2\Gamma^2}{\Gamma^2+\left(\frac{\Dm}{2}\right)^2}\frac{\Re e(\omega)}{1+\mathcal F(\abs{\omega}^2)} ~~,\label{eq:Asymmetries:04b}
\end{equation}
where $\mathcal F(\abs{\omega}^2)=\frac{1}{2}\abs{\omega}^2\frac{2\Gamma^2+(\Dm)^2}{\Gamma^2+(\Dm)^2}$.

This concludes our analysis on the $\omega$-effect-induced demise of flavour 
tagging in $B$-meson factories. We stress once more that 
the above-described effects are specific to a particular kind of CPT violation
invoking decoherence, which affects the {\it identity } of the (initial) 
neutral meson states~\cite{Bernabeu:2003ym}, and is in principle unrelated to 
the dynamics of their evolution.
This is clearly distinguishable from  other types of 
CPT violation existing in the literature, e.g. 
those pertaining to 
the non-commutativity of the CPT operator with 
the matter hamiltonian~\cite{kostelecky}, or those related to 
non-local field theory
models~\cite{local}, 
or even those associated with a decoherent temporal {\it evolution} of 
matter in quantum gravity media~\cite{lopez}.  
It is 
hoped that studies in $B$-factories such as the one suggested above
will improve the bounds 
on such effects significantly in the foreseeable future. 
Together with other neutral meson factories, such as 
$\phi$-factories~\cite{Bernabeu:2003ym}, this system
may then provide essential probes for novel physics, associated 
with effects of quantum gravity on entangled states.
This should be viewed as complementary to other quantum gravity 
studies.

{\bf Acknowledgments}

The authors thank F.J.Botella for useful discussions. 
This work has been supported 
 by the Grant CICYT FPA2002-00612. 
E.A is indebted 
to the University of Valencia for financial support. 
M.N acknowledges Ministerio de Educaci\'on y Ciencia for his FPU grant.
N.E.M thanks the Theoretical Physics Department
of the University of Valencia for the hospitality extended to him 
during this collaboration.


\begin{thebibliography}{99}

\bibitem{tagging} see for instance: R.~J.~Wilson,
``Neutral B meson flavor tagging,'',
{\it Prepared for 5th International Linear Collider Workshop (LCWS 2000), Fermilab, Batavia, Illinois, 24-28 Oct 2000}, and references therein;
see also: H.~Kakuno {\it et al.},
arXiv:hep-ex/0403022.

\bibitem{Abe}
K.~Abe {\it et al.}  [BELLE Collaboration],
arXiv:hep-ex/0408111.

\bibitem{dunietz} H.~J.~Lipkin,
Phys.\ Rev. {\bf 176}, 1715 (1968);
Phys.\ Lett.\ B {\bf 219}, 474 (1989);
I.~Dunietz, J.~Hauser and J.~L.~Rosner,
Phys.\ Rev.\ D {\bf 35}, 2166 (1987);


\bibitem{botella} 
J.~Bernabeu, F.~J.~Botella and J.~Roldan,
Phys.\ Lett.\ B {\bf 211}, 226 (1988);
{\it Proc. 25th Anniversary of the Discovery of CP Violation, Blois, France, May 22-26, 1989}, 
Blois CP Violations 1989:0389-400 (QCD161:I44:1989).  

\bibitem{bernabeu} L.~Wolfenstein, Nucl. Phys. {\bf B246}, 45 (1984);
M.~C.~Banuls and J.~Bernabeu,
Phys.\ Lett.\ B {\bf 464}, 117 (1999);
Nucl.\ Phys.\ B {\bf 590}, 19 (2000).

\bibitem{foam} see for instance: 
J.A. Wheeler and K. Ford, {\it Geons, Black Holes
and Quantum Foam: A life in Physics} (New York, USA: Norton (1998)); 
S. Hawking, Comm. Math. Phys. {\bf 43}, 199 (1975);
{\it ibid.} {\bf 87}, 395 (1982); J. Bekenstein, Phys. Rev. {\bf D12}, 
3077 (1975); J.L. Freedman and R. Sorkin, Phys. Rev. Lett. {\bf 44},
1100 (1980); Gen. Rel. Grav. {\bf 14}, 615 (1982);
J. Ellis, J. Hagelin, D.V. Nanopoulos and M. Srednicki,
Nucl. Phys. {\bf B241}, 381 (1984).



\bibitem{Bernabeu:2003ym}
J.~Bernabeu, N.~E.~Mavromatos and J.~Papavassiliou,
Phys.\ Rev.\ Lett.\  {\bf 92}, 131601 (2004).


\bibitem{cpt} R.F. Streater and A.S. Wightman, {\it PCT, Spin and Statistics, and All That}, p. 207 (Redwood City, USA: Addison Wesley (1989)).

\bibitem{wald} R. Wald, Phys. Rev. {\bf D21}, 2742 (1980). 

\bibitem{CPTDeltaBQ}
G.~V.~Dass and K.~V.~L.~Sarma,
Phys.\ Rev.\ D {\bf 54} 5880 (1996);
Eur.\ Phys.\ J.\ C {\bf 5} 283 (1998);
Z.~z.~L.~Xing,
Phys.\ Lett.\ B {\bf 450}, 202 (1999);
L.~Lavoura and J.~P.~Silva,
Phys.\ Rev.\ D {\bf 60}, 056003 (1999);
G.~V.~Dass, W.~Grimus and L.~Lavoura,
JHEP {\bf 0102}, 044 (2001);
K.~R.~S.~Balaji, W.~Horn and E.~A.~Paschos,
Phys.\ Rev.\ D {\bf 68}, 076004 (2003).

\bibitem{ASemileptonic}
G.~Abbiendi {\it et al.}  [OPAL Collaboration],
Eur.\ Phys.\ J.\ C {\bf 12}, 609 (2000);
D.~E.~Jaffe {\it et al.}  [CLEO Collaboration],
Phys.\ Rev.\ Lett.\  {\bf 86}, 5000 (2001);
R.~Barate {\it et al.}  [ALEPH Collaboration],
Eur.\ Phys.\ J.\ C {\bf 20} (2001) 431;
B.~Aubert {\it et al.}  [BABAR Collaboration],
Phys.\ Rev.\ Lett.\  {\bf 88}, 231801 (2002).


\bibitem{kostelecky} see for instance: V.~A.~Kostelecky,
arXiv:hep-ph/0104227, and references therein.

\bibitem{local} G.~Barenboim and J.~Lykken,
Phys.\ Lett.\ B {\bf 554}, 73 (2003)
[arXiv:hep-ph/0210411].


\bibitem{lopez} J.~R.~Ellis, N.~E.~Mavromatos and D.~V.~Nanopoulos,
Phys.\ Lett.\ B {\bf 293}, 142 (1992);
 Int.\ J.\ Mod.\ Phys.\ A {\bf 11} (1996) 1489
[arXiv:hep-th/9212057]; 
J.~R.~Ellis, J.~L.~Lopez, N.~E.~Mavromatos and D.~V.~Nanopoulos,
Phys.\ Rev.\ D {\bf 53}, 3846 (1996); 
P.~Huet and M.~E.~Peskin,
Nucl.\ Phys.\ B {\bf 434}, 3 (1995); 
F.~Benatti and R.~Floreanini,
Phys.\ Lett.\ B {\bf 468} (1999) 287.

\bibitem{foot}
Different analyses of CPT violation and  
$\Delta B\neq\Delta Q$ can be found in \cite{CPTDeltaBQ}.  



\end{thebibliography}
\end{document}